\documentclass{ws-procs10x7}

\begin{document}

\title{Ultra--High-Energy Cosmic Rays}

\author{Stefan Westerhoff}

\address{Columbia University, Department of Physics, New York, NY 10027,
USA\\E-mail: westerhoff@nevis.columbia.edu}

\twocolumn[\maketitle\abstract{
One of the most striking astrophysical phenomena today is the
existence of cosmic ray particles with energies in excess of
$10^{20}$\,eV.  While their presence has been confirmed by a number
of experiments, it is not clear where and how these particles
are accelerated to these energies and how they travel astronomical
distances without substantial energy loss.  
We are entering an exciting new era in cosmic ray physics, with 
instruments now producing data of unprecedented quality and quantity 
to tackle the many open questions.
This paper reviews the current experimental status of cosmic ray physics
and summarizes recent results on the energy spectrum
and arrival directions of ultra-high-energy cosmic rays.}]

\section{Introduction}

Cosmic ray particles were discovered almost one hundred years
ago, and yet very little is known about the origin of the most energetic
particles above and around $10^{18}$\,eV, traditionally referred to as
``ultra--high-energy cosmic rays.''
The measured spectrum of cosmic rays extends beyond $10^{20}$\,eV, 11 orders
of magnitude greater than the equivalent rest mass of the proton.  While
the presence of particles at these energies has been confirmed by a number of 
experiments, it is not clear where and how these particles are accelerated to 
these energies and how they travel astronomical distances without substantial
energy loss.  Some indication comes from the energy spectrum itself, which
roughly follows a power law $E^{-2.8}$ and is therefore ``non-thermal.'' 
The power law behavior and the ``universal'' spectral index indicate 
that the underlying acceleration mechanism could be Fermi acceleration\cite{fermi} 
at shock fronts, e.g. in Supernova Remnants (SNRs) and in the jets of Active 
Galactic Nuclei (AGNs).  Regardless of the actual acceleration process, 
it is clear that thermal emission processes cannot generate such energies.

Several quantities accessible to experiment can help to reveal the sources 
of ultra--high-energy cosmic rays: the {\it flux} of cosmic rays; their 
{\it chemical composition}; and their {\it arrival direction}. Charged cosmic ray 
primaries are subject to deflection in Galactic and intergalactic 
magnetic fields and do not necessarily point back to their sources.  The strength 
and orientation of these fields is poorly known and estimates 
vary\cite{sigl2003,dolag2003}, but their impact should decrease at the highest 
energies, above several times $10^{19}$\,eV; here, cosmic ray astronomy might
be possible.

Since the cosmic ray flux drops almost three orders of magnitude for each energy
decade, the flux at the highest energies is very low, about one particle per
$\mathrm{km}^{2}$ per year above $5\times 10^{18}$\,eV.  Low statistics have 
historically plagued the field; the total published number of events above 
$4.0\times 10^{19}$\,eV is still less than 100, and drawing conclusions on the basis 
of such a small data set has proved rather perilous.  
However, the experimental situation is finally improving;  new instruments
are now collecting data of unprecedented quality and quantity.  Since 1999,
the High Resolution Fly's Eye (HiRes) air fluorescence stereo 
detector in Utah has accumulated data with excellent angular resolution. 
The HiRes data set has been used extensively over the last two years to search for 
small-scale anisotropies in the arrival directions of cosmic rays,
correlations of cosmic rays with known astrophysical sources, and to study 
the composition of the primary cosmic ray flux.

In the southern hemisphere, the world's largest detector for cosmic radiation,
the Pierre Auger Observatory in Argentina, is nearing completion, and first results
on the energy spectrum and the arrival direction distribution of cosmic
rays have recently been published.  

In this paper, I will review some recent developments in ultra--high-energy cosmic 
ray physics.  The paper is organized as follows: Section 2 gives a short review of the
experimental techniques and the current major instruments in the field.  I will then discuss new 
results from the Auger and HiRes experiments on the energy spectrum (Section 3)
and the arrival direction distribution (Section 4).  Concluding remarks follow in
Section 5.

\begin{figure}
\epsfxsize180pt
\figurebox{120pt}{160pt}{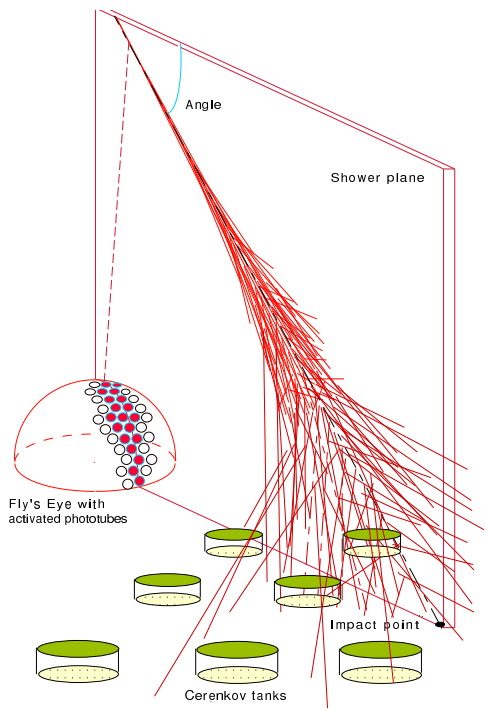}
\caption{Scheme of the different cosmic ray detection techniques (Auger Collaboration).}
\label{fig:hybrid}
\end{figure}

\section{Experimental Techniques}

Because the flux at ultra--high energies is small, experiments need a large detector
volume.  Consequently, detectors have to be earth-bound, and the primary cosmic ray 
particles can not be observed directly.  Primaries interact in the upper atmosphere
and induce extensive air showers with on the order of $10^{10}$ particles for a 
$10^{19}$\,eV primary.  The properties of the original cosmic ray particle, such as 
arrival direction and energy, have to be inferred from the observed properties of
the extensive air shower.

\begin{table*}[t]
\caption{Comparison of Instruments. {\it Comments:} $^{1)}$ above $10^{20}$\,eV, $10\,\%$ duty cycle,
$^{2)}$ January 2004 -- June 2004, $^{3)}$ error bars are strongly asymmetric \label{tab:exp}}
\begin{tabular}{|c|c|c|c|c|c|}  
\hline
Experiment & Operation & Aperture & Exposure & Angular Resolution & $\Delta E/E$ \\
           &           & $\left[ \mathrm{km}^{2}\,\mathrm{sr} \right]$
                       & $\left[ \mathrm{km}^{2}\,\mathrm{sr}\,\mathrm{yr} \right]$ 
                       & (68\,\%)
                       & $\left[\%\right]$ \\
\hline
\hline
AGASA             & 1984-2003 & $\simeq 250$ & 1620      & $2.5^{\circ}$  & 25 \\
\hline
HiRes mono        & 1997 -    & $10000^{1)}$ & 5000      & $>2.5^{\circ, 3)}$ & 25 \\
HiRes stereo      & 2000 -    & $10000^{1)}$ & 3400      & $0.6^{\circ}$  & 15 \\
\hline
Auger (under      & 2004 -    & 7400         & 1750 (SD)$^{2)}$ & $2.0^{\circ}$ - $0.9^{\circ}$ (SD)  & 10 \\
construction)     &           &              &           & $0.6^{\circ}$                 (FD)  &    \\
\hline
\end{tabular}
\end{table*}

There are two different techniques to study cosmic ray air showers at ultra--high-energies.
Both are shown schematically in Fig.\,\ref{fig:hybrid}.
{\it Ground arrays} sample the shower front arriving on the Earth's surface with an array of
particle detectors, for example scintillation counters or water Cherenkov detectors.  The 
arrival direction of the air shower and the primary cosmic ray particle is reconstructed from 
the differences in trigger times for individual detectors as the narrow shower front passes.  
The advantage of ground arrays is their near $100\,\%$ duty cycle and the robustness of the 
detectors.  A disadvantage is that ground arrays sample the shower at one altitude only and do not 
record the development of the shower in the atmosphere.  Moreover, the sampling density is 
typically very small.

The classic example of a pure ground array is the AGASA (Akeno Giant Air Shower Array) 
experiment, which operated in Japan from 1984 to 2003.  In its final stage, 
the array consisted of 111 scintillation counters of $2.2\,\mathrm{m}^{2}$ area each on a 1\,km 
spacing, leading to a total area of about $100\,\mathrm{km}^{2}$.

Apart from the cascade of secondary particles, air showers also produce Cherenkov light and 
fluorescence light.  The latter is produced when the particles in the air shower cascade
excite air molecules, which fluoresce in the UV.  {\it Air fluorescence detectors} measure this
light with photomultiplier cameras that observe the night sky.  The shower is observed by a 
succession of tubes and reconstructed using the photomultiplier timing and pulseheight
information.  Air fluorescence detectors can only operate on clear, moonless nights
with good atmospheric conditions, so the duty cycle is about $10\,\%$; however,
they observe the shower development in the atmosphere and provide us with a nearly calorimetric 
energy estimate.  In addition, the instantaneous detector volume is rather large, of order 
$10000\,\mathrm{km}^{2}\,\mathrm{sr}$ at $10^{20}$\,eV.

The High Resolution Fly's Eye (HiRes) experiment in Utah is a stereo air fluorescence
detector with 2 sites roughly 13\,km apart.  Each site is made up of several telescope 
units monitoring different parts of the night sky.  With 22 (42) telescopes at the first 
(second) site, the full detector covers about $360^{\circ}$ ($336^{\circ}$) in azimuth 
and $3^{\circ}-16.5^{\circ}$ ($3^{\circ}-30^{\circ}$) in elevation above horizon.  Each 
telescope consists of a mirror with an area of about $5\,\mathrm{m}^{2}$ for 
light collection and a cluster of 256 hexagonal photomultiplier tubes in the focal plane.

The HiRes air fluorescence detector can operate in ``monocular mode,'' with air 
showers only observed from one site, or in ``stereoscopic mode,'' with the same shower
observed by both detectors simultaneously.  The monocular operation suffers from poor 
angular and energy resolution.  With only one ``eye,'' the shower-detector-plane
({\it i.e.} the plane that contains the shower and the detector (see Fig.\,\ref{fig:hybrid}))
can be reconstructed with high accuracy.  Unfortunately, the position of the shower within that
plane, determined using the photomultiplier trigger times, is ambiguous.  Stereo 
viewing of the shower with two detectors breaks the ambiguity and leads to an excellent 
angular resolution of order $0.5^{\circ}$.  HiRes has been taking data in monocular mode 
since 1996 and in stereo mode since December 1999.

In summary, stereo air fluorescence detectors have excellent angular resolution and give a nearly
calorimetric energy determination, while ground arrays have the advantage of a relatively
straightforward determination of the detector aperture.  Obviously, the best detector
is a detector that combines the two techniques.
The Pierre Auger Observatory, currently under construction in Malargue, Argentina,
and scheduled to be completed in 2006, is such a {\it hybrid detector}, combining
both a ground array and fluorescence detectors.  
  
The Auger surface detector array\cite{auger_sd} will eventually comprise 1600 detector 
stations with 1500\,m separation, spread over a total area of $3000\,\mathrm{km}^{2}$.  
One year of Auger data-taking will therefore correspond to about 30 AGASA years.  Each 
detector station is a light-tight 11\,000 liter tank filled with pure water.  Three 9-inch 
photomultipliers in the tank measure the Cherenkov light from shower particles crossing 
the tank.  The stations are self-contained and work on solar power.  The surface detector 
array is complemented by 4 fluorescence stations\cite{auger_fd} with 6 telescopes each.  
The field of view of a single telescope covers $30^{\circ}$ in azimuth and $28.6^{\circ}$ 
in elevation, adding up to a total field of view of $180^{\circ}\times 28.6^{\circ}$ for 
each site.  Each telescope consists of a spherical mirror of area
$3.5\,\mathrm{m}\times 3.5\,\mathrm{m}$ and a focal surface with 440 hexagonal (45\,mm diameter) 
photomultipliers.  Because of the large field of view of each telescope, a Schmidt optics
is used.

While still under construction, the Auger experiment has recorded data with a growing
detector since January 2004.  The total exposure of the surface detector array already 
exceeds the total AGASA exposure, making Auger a competitive experiment.  

Table\,1 gives an overview of experimental parameters for AGASA, Auger, HiRes monocular 
and HiRes stereo, including the aperture of the instruments and the exposure used in
publications of recent results.

\section{Results}

Until a few years ago, the world data set of ultra--high-energy cosmic rays was
dominated by data recorded with the AGASA air shower array.  In a number of important
publications, the AGASA group has described several exciting and controversial results,
including the shape of the energy spectrum above $10^{18}$\,eV, studies of possible
anisotropies in the arrival directions above $4.0\times 10^{19}$\,eV, and an excess
of cosmic ray flux from the Galactic center region around $10^{18}$\,eV.
Due to the small number of events, some of these results have a small statistical
significance.  For several years, statistically independent data sets to support or
refute these findings were not available.  Only recently, with the start of
HiRes stereo data-taking in 1999 and the beginning of operations at the Pierre Auger
Observatory in January 2004, are we reaching the point where we can study these topics
with new data.

\subsection{Energy Spectrum}

Proton primaries above $5.0\times 10^{19}$\,eV interact with the 2.7\,K microwave 
background via photo-pion production, losing energy in each interaction until they 
eventually fall below the energy threshold.  This so-called GZK effect,          
postulated by Greisen\cite{gzk1} and independently by Zatsepin and Kuzmin\cite{gzk2} 
shortly after the discovery of the microwave background, is expected to lead to a 
rapid fall-off of the cosmic ray energy spectrum above this energy, but it has    
not been experimentally confirmed at this point.

The AGASA group claimed\cite{agasa_energy1,agasa_energy2} that the GZK suppression 
is not observed, raising questions as to the nature of the primary particles or even the 
particle physics involved.  More recently, the HiRes monocular 
spectrum\cite{hires1_spectrum,hires2_spectrum} has been interpreted as being in agreement 
with a GZK suppression.  While the disagreement between the AGASA and the HiRes mono result
has received a lot of attention, it has also been pointed out that the two spectra actually 
agree reasonable well if systematic errors are taken seriously\cite{olinto,waxman}.  

\begin{figure*}
\epsfxsize250pt
\figurebox{120pt}{160pt}{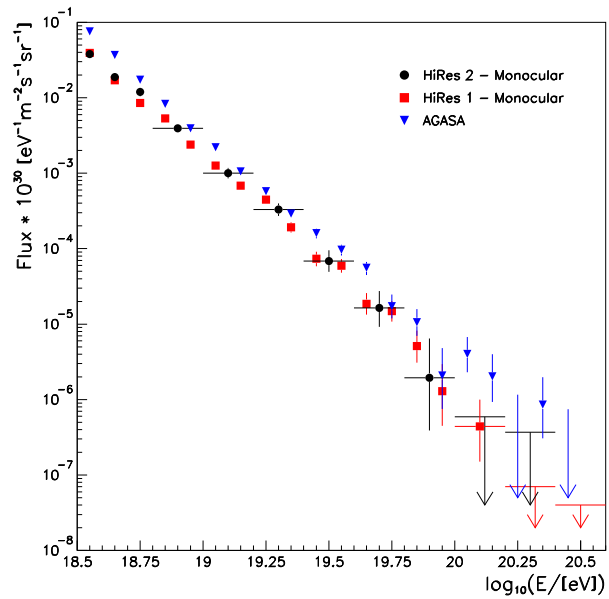}
\caption{Ultra--high-energy cosmic ray flux as a function of energy as measured by the AGASA 
experiment$^{9}$ and the HiRes detectors in monocular mode, 
HiRes\,1$^{10}$ and HiRes\,2$^{11}$.}
\label{fig:spectrum_agasa_hires}
\end{figure*}

Fig.\,\ref{fig:spectrum_agasa_hires} depicts the situation.  It shows the differential energy 
spectrum as reported by AGASA\cite{agasa_energy2} and HiRes\cite{hires1_spectrum,hires2_spectrum} 
in monocular mode (the HiRes collaboration has not yet published a stereo spectrum).  While only 
statistical errors are shown, systematic errors of order 25\,\% (roughly what is reported by 
the experiments) would manifest themselves simply by a shift of one bin in this plot.  Keeping 
this in mind, both spectra agree quite well below $10^{19.8}$\,eV.  The disagreement at the 
highest energies has been estimated to be of order $2.0\,\sigma$\,\cite{olinto}.

Reversing the argument, it is actually quite surprising how well the spectra agree,
given that ground arrays and fluorescence detectors have entirely different systematic 
errors in their estimate of the shower energy.  AGASA has given a detailed account of 
their energy resolution\cite{agasa_energy2}.  Ground arrays determine the energy of 
a shower from the measured particle density $S$ at some (fixed) distance from the shower 
core\cite{hillas}.  The optimal distance is the one where the spread in particle density 
is minimal, and this optimal distance depends on the detector geometry (600\,m for 
AGASA\cite{agasa_energy2} and 1000\,m for Auger\cite{auger_s1000}).  The correlation between 
the particle density and the energy is then established using simulated data.  This of
course leads to a strong mass and model dependence.

Air fluorescence detectors image the shower development in the atmosphere and obtain a 
nearly calorimetric signal.  However, the Earth's atmosphere is a tricky calorimeter. 
It is not only highly inhomogeneous, but also changes on short time scales.  The atmosphere
needs to be monitored continuously in order to be able to correctly account for Rayleigh
and Mie scattering.  In addition, there are systematic errors from the
fluorescence yield, the absolute calibration, and the accounting for ``missing energy,''
{\it i.e.} energy that goes into particles that do not contribute to air fluorescence.

\begin{figure*}
\epsfxsize250pt
\figurebox{120pt}{160pt}{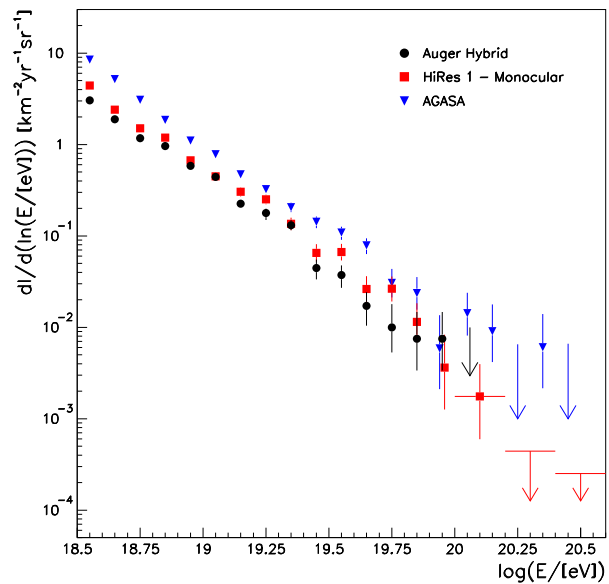}
\caption{Intensity of the ultra--high-energy cosmic ray flux as a function of energy measured 
by the AGASA experiment$^{9}$ and HiRes\,1 in monocular mode$^{10}$, 
compared to a first estimate by the Pierre Auger Observatory$^{14}$.}
\label{fig:spectrum_all}
\end{figure*}

The hybrid nature of the Pierre Auger Observatory will enable us to understand whether 
there are intrinsic difficulties with one of the two methods that contribute to the 
disagreement in the two spectra.  At this time, the Auger experiment has not yet accumulated 
the necessary statistics to make a definitive statement on the shape of the energy 
spectrum at the highest energies.  The collaboration has, however, already published a first 
estimate of the spectrum.  Fig.\,\ref{fig:spectrum_all} compares the Auger spectrum to both 
the AGASA and HiRes\,1 monocular measurement.  The spectrum is based on 3525 events with
energies above $3\times 10^{18}$\,eV data taken with the ground array between
January 2004 and June 2005.  At these energies, the ground array is fully efficient and the
total exposure can be inferred from the total running time (where the growing size of the
detector is accounted for) and the geometrical aperture of the ground array.
The energy of the showers is established using the subset of {\it hybrid} events
in the data sample, {\it i.e.} the subset of events that are also detected in one of the
fluorescence detectors.  These events are used to derive a relationship between shower 
energy (as determined by the fluorescence detector) and the ground parameter $S$ (as measured
by the ground array)\cite{auger_spectrum}.  

By using the statistics and the exposure of the ground array and the nearly calorimetric
energy estimate of the fluorescence detectors, it is ensured that the Auger
spectrum does not rely strongly on either simulations or assumptions about the chemical 
composition of the cosmic ray flux.  Unlike the AGASA and HiRes spectra, it is therefore
nearly model-independent. 

As shown in Fig.\,\ref{fig:spectrum_all}, the Auger spectrum agrees quite well with the HiRes 
measurement at this point.  Considering the large systematic uncertainties, there is also 
little disagreement with AGASA.  For this first estimate of the spectrum, systematic errors range 
from about 30\,\% at $3\times 10^{18}$\,eV to 50\,\% at $10^{20}$\,eV, of which about 25\,\% stem 
from total systematic uncertainties in the fluorescence detector energy measurements.  
Errors will soon become considerably smaller with larger statistics and the application of
more sophisticated reconstruction methods.

Two additional aspects of the Auger spectrum should be stressed.

(1) While the spectrum does not show any events above $10^{20}$\,eV, the Auger experiment
has detected an event above that energy\cite{auger_events}.  The event does not enter the
energy spectrum because its core falls outside the surface detector array, but the
event is well-reconstructed and passes all other quality cuts.  This means that all 
major cosmic ray experiments have now confirmed the existence of particles above $10^{20}$\,eV.

(2) As described above, the energy scale of the spectrum is normalized to the fluorescence 
detector.  When simulations are used to determine the shower energy from surface detector data 
alone, the energies are systematically higher by at least 25\,\%.  Energies determined in this way
depend on the shower simulation code, the hadronic model, and the assumed composition of the 
cosmic ray flux.  The difference shows that many systematics need to be addressed.
Auger is in a unique position to study these systematic uncertainties and will measure the 
spectrum in the southern hemisphere accurately in coming years.  There is no reason to believe 
that the energy spectra in the northern sky and the southern sky are identical, although 
differences are probably subtle and will require a larger instrument in the northern hemisphere.

\subsection{Search for Small-Scale Clustering and Point Sources}

A direct way to search for the sources of ultra--high energy cosmic rays is to study
their arrival direction distribution.  Astronomy with charged particles of course faces
a serious problem.  At low energies, Galactic and intergalactic magnetic fields will 
bend the particle's trajectories sufficiently to render the direction information useless.
However, the Larmor radius increases with energy, and cosmic ray astronomy may be possible
above some energy threshold.  But little is known about intergalactic magnetic fields, so 
it is not straightforward to determine an optimal energy cut for arrival direction studies.
Choosing an energy threshold too low means that deflections destroy any correlation, but too
high a threshold also weakens the statistical power of the data set.  Given these uncertainties, 
an {\it a priori} optimal choice for the energy threshold and the angular separation for 
clustering searches does not exist.

AGASA claimed a significant amount of clustering in the arrival direction 
distribution as early as 1996\cite{agasa_cluster96}  and has updated these results 
frequently\cite{agasa_cluster99,agasa_cluster00,agasa_cluster01,agasa_cluster03}.  
The current claim is that the 5 ``doublets'' and 1 ``triplet'' observed in the 
set of 57 events with energies above $4.0\times 10^{19}$\,eV have a probability of 
less than 0.1\,\% to occur by chance in an AGASA data set of this size.  This is an 
extremely important result, but its validity has been questioned on statistical 
grounds.  The problems stem from the fact that the data set used to 
evaluate the chance probability {\it includes} the data used for formulating the 
clustering hypothesis in the first place; parameters like the energy threshold 
and the angular scale that defines a ``cluster'' of arrival directions were not 
defined {\it a priori}. The probability of 0.1\,\% has therefore little meaning, 
and analyses that test the hypothesis only with the part of the data set that was 
recorded after the hypothesis was formulated find a much higher chance probability 
of around 8\,\%\,\cite{finley2003}, making the result insignificant.

The AGASA clustering claim has been tested with the statistically independent HiRes
stereo data set.  The HiRes collaboration has published several papers describing 
searches for deviation from isotropy, including the calculation of the standard 
two-point correlation function\cite{finley_dpf2004}, a two-point correlation 
scan\cite{scan2004}, and a search for point sources using an unbinned maximum 
likelihood ratio test\cite{likeli2005,likeli_icrc}.  None of these searches have 
uncovered any deviation from isotropy, and no evidence for statistically significant 
point sources in the HiRes or the combined HiRes/AGASA data sets above 
$4.0\times 10^{19}$\,eV or $1.0\times 10^{19}$\,eV were found\cite{likeli2005,likeli_icrc} 
(although a point source has been claimed by other 
authors\cite{glennys2005a,glennys2005b}).  With an overall exposure that exceeds the 
AGASA exposure and an angular resolution that is 4 to 10 times sharper than AGASA's, 
HiRes stereo finds no small-scale anisotropy at any energy threshold above 
$10^{19}$\,eV and any angular scale out to $5^{\circ}$.

\subsection{Correlation with BL Lac Objects}

There has recently been marginal evidence that a small fraction of the ultra--high-energy
cosmic ray flux originates from BL Lacertae Objects\cite{gorbunov2004,bllac2005}.  
BL Lacs are a subclass of blazars, which are active galaxies in which the jet axis 
happens to point almost directly along the line of sight.  The EGRET instrument on board 
the Compton Gamma Ray Observatory (CGRO) has firmly established blazars as sources of 
high energy $\gamma$-rays above 100\,MeV\,\cite{hartman1999}, and several BL Lac objects 
have been observed at TeV energies with ground-based air Cherenkov telescopes\cite{horan2004}.  
High energy $\gamma$-rays could be by-products of electromagnetic cascades from energy
losses associated with the acceleration of ultra--high-energy cosmic
rays and their propagation in intergalactic space\cite{berezinskii1990,coppi1997}.

The history of claims for a correlation between BL Lacs and ultra--high-energy cosmic rays
is rather convoluted.  Correlation claims based on data recorded with AGASA and the Yakutsk 
array have been published since 2001\cite{tt2001,tt2002,gorbunov2002}, but a problematic 
aspect of the claims is the procedure used to establish correlations and evaluate their 
statistical significance.  The authors explicitly tuned their selection criteria to assemble 
catalogs showing a maximum correlation with arrival directions of cosmic rays above some 
energy.  The statistical significance quoted is therefore meaningless, and claims of BL Lac 
correlations have been criticized on these grounds\cite{evans2003,stern2005}.  In some cases 
it has been shown that statistically independent data sets do not confirm the 
correlations\cite{torres2003}.

Recently, the HiRes collaboration published its own analysis of the data\cite{bllac2005},  
using an unbinned maximum likelihood ratio test which accounts for the individual point 
spread function of each event.  The HiRes stereo data does not confirm any of the previous 
claims based on AGASA and Yakutsk data, in spite of its larger statistical power.  It does, 
however, verify a recent analysis\cite{gorbunov2004} of correlations between published HiRes 
stereo events above $10^{19}$\,eV and a subset of confirmed BL Lacs from the 10th Veron 
Catalog\cite{veron}.  This subset\cite{tt2001} contains 157 confirmed BL Lacs from the catalog 
with optical magnitude $m<18$.  Since the cuts\cite{gorbunov2004} used to isolate this signal 
are not {\it a priori}, the correlation needs to be confirmed with statistically independent 
data before any claims can be made.  

\begin{figure*}
\epsfxsize320pt
\figurebox{120pt}{160pt}{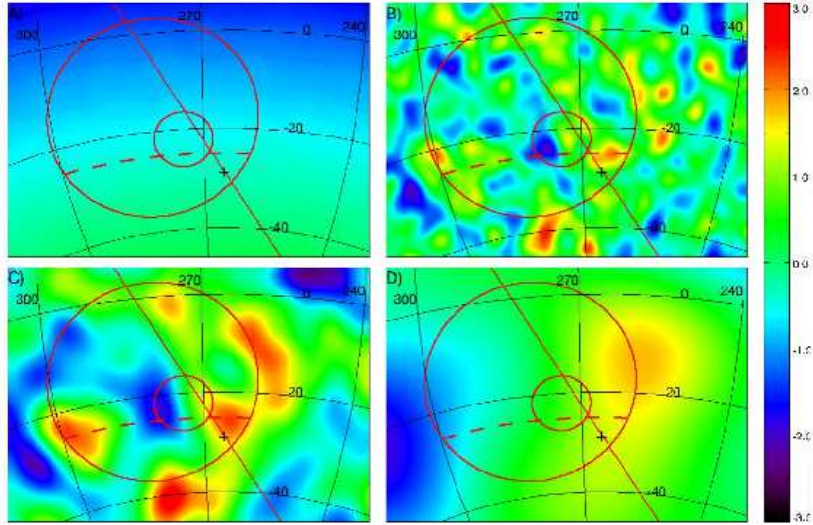}
\caption{Auger skymap in  the vicinity of the Galactic center$^{49}$, using
surface detector data.  Shown are the coverage map (upper left) and the significance maps 
for the data smoothed using three different resolutions: the Auger 
resolution (upper right), the large integration radius used by the AGASA collaboration 
for their claim of a Galactic center source (lower right), and the resolution of 
the SUGAR experiment (lower left).  The Galactic center is indicated by a cross, the regions 
of excess for AGASA and SUGAR are indicated by the red circles.  The dashed line indicates the 
limit of the AGASA field of view.}
\label{fig:galactic_center}
\end{figure*}

The correlations are strongest at energies around $10^{19}$\,eV, where magnetic fields 
are expected to sufficiently scramble the arrival directions of charged particles, yet 
the correlations are on the scale of the detector angular resolution.  This would suggest 
neutral cosmic ray primaries for these events, or at least that the primaries were neutral
during significant portions of their journey through Galactic and extragalactic magnetic 
fields.  If verified with future data, the correlation with BL Lac objects would be the 
first evidence for an extragalactic origin of the highest energy cosmic rays, and a
first indication that at least a fraction of this flux originates in known astrophysical 
objects.  

\subsection{Galactic Center}

One of the most interesting regions of the sky to search for an excess of ultra--high-energy
cosmic rays is around the Galactic center.  This region is a natural site for cosmic ray
acceleration.  It harbors a black hole of mass $2.6\times 10^{6}$ solar masses whose position
is consistent with the radio source Sagittarius $\mathrm{A}^{\star}$.  Hour-scale X-ray and 
rapid IR flaring indicate the presence of an active nucleus with low bolometric luminosity. 
In addition, this crowded part of the sky contains a dense cluster of stars, stellar remnants,
and the supernova remnant Sgr A East.  The Galactic center is now also established as a source
of TeV $\gamma$-radiation\cite{cangaroo,whipple,hess}.

The AGASA experiment reported an excess with a statistical significance of order $4\,\sigma$ 
near the Galactic center\cite{agasa_galaxy1,agasa_galaxy2}.  The excess is observed in a narrow 
energy band from $10^{17.9}$\,eV to $10^{18.3}$\,eV only.  Due to its location in 
the northern hemisphere and a zenith angle cut of $60^{\circ}$, AGASA's field of view cuts off 
roughly $5^{\circ}$ north of the Galactic center, so the center itself, at right ascension 
$\alpha=266.4^{\circ}$ and declination $\delta=-28.9^{\circ}$, is outside the field of view.
The AGASA excess is found roughly around $\alpha=280^{\circ}$, $\delta=-16^{\circ}$, so it is
offset considerably from the location of the Galactic center.  To produce this excess, the event 
density is integrated over a circle with a radius of $20^{\circ}$.  Several other ``beam sizes'' 
were tried, but this integration radius was found to maximize the signal. 

The chance probability in the AGASA publication is {\it a posteriori}.  Since the analysis has 
several ``tunable'' parameters like the energy bin and the integration radius, the true chance 
probability cannot be derived from the AGASA data set itself.  The AGASA result was, however, 
supported by a re-analysis of archival data taken with the SUGAR array\cite{sugar}, a cosmic ray 
experiment that operated in Australia between 1968 and 1979.  Unlike AGASA, SUGAR operated 
from a location with good visibility of the Galactic center region.  The SUGAR data showed an
excess of about $2.9\,\sigma$ in the energy bin $10^{17.9}$\,eV to $10^{18.5}$\,eV, roughly the 
same energy bin as AGASA, but at $\alpha=274^{\circ}$ and $\delta=-22^{\circ}$, so offset both
from the location of the AGASA excess and the Galactic center.  Furthermore, the SUGAR excess
was consistent with a point source, indicating neutral primaries.

In such a context, it is interesting to point out that {\it neutron} primaries are a viable 
hypothesis, as they could travel undeflected.  The neutron hypothesis would also explain 
the narrow energy range of the signal.  In an amazing coincidence, the neutron decay length 
at $10^{18}$\,eV roughly corresponds to the distance between us and the Galactic center, 
about 8.5\,kpc.

Like SUGAR, the Auger detector has good visibility of the Galactic center region.  
The Auger coverage map in the vicinity of the Galactic center is smooth and shows 
no strong variations (see upper left part of Fig.\,\ref{fig:galactic_center}).
The Auger group has searched for an excess in the Galactic center region in the data set 
taken from 1 January 2004 through 5 June 2005\cite{auger_galaxy}, both in the surface detector
data (angular resolution $1.5^{\circ}$) and hybrid data ($0.6^{\circ}$).  The event statistics 
for the surface detector data are already larger than those of the two previous experiments (3 times 
AGASA, 10 times SUGAR), but the AGASA excess occurs in an energy range where Auger is not fully 
efficient ($>30\,\%$ for protons, $>50\,\%$ for iron).  However, even in the worst case (a proton 
signal on an iron-dominated background) a $5.2\,\sigma$ excess is expected if the AGASA source 
is real.

Fig.\,\ref{fig:galactic_center} shows the results of an analysis using surface detector data
taken between January 2004 and June 2005.  The figure shows the map smoothed using different 
angular resolutions, corresponding to the Auger resolution (upper right), the AGASA signal
(lower right), and the SUGAR resolution (lower left).  No significant excess is found in all
cases.  The Auger group also studied neighboring energy bins to account for a possible systematic
difference in the energy calibration, but no excess is found in any scenario.
For the Galactic center itself, a 95\,\% confidence level upper limit for the flux from a point source
is derived\cite{auger_galaxy} which excludes the neutron source scenario suggested to explain the
previous claims\cite{agasa_galaxy2,bossa}.

\section{Concluding Remarks}
 
The most important and most fascinating result at this point is that {\it cosmic ray
particles with energies above $10^{20}$\,eV exist} -- their existence has been confirmed 
by all experiments, regardless of the experimental technique, and regardless of whether the
GZK suppression is present or absent in the data.  

With current statistics, the cosmic ray sky is remarkably isotropic.  One possible reason for 
this could be magnetic smearing.  This effect should be smallest at the highest energies, 
above $10^{20}$\,eV, but until now the number of events has been too small to draw any                
conclusions.  In another possible scenario, we could be dealing with many sources, each 
currently contributing at most one or two events, and again, more data will help to eventually 
resolve the strongest ones.

The Pierre Auger Observatory will dramatically increase the high-energy sample size over the 
next several years.  Auger is still under construction (to be completed in 2006), but the 
collaboration has already published first results, among them a first estimate of the cosmic ray energy
spectrum.  With its location in the southern hemisphere, its hybrid design and unprecedented 
size, Auger is in an excellent position to answer definitively the questions left open by the
previous generation of cosmic ray experiments, and to make the discoveries that will challenge 
the next one.

\section*{Acknowledgments}
It is a pleasure to thank the organizers of Lepton-Photon 2005 for a delightful meeting
in Uppsala.
The author is supported by the National Science Foundation under
contract numbers NSF-PHY-9321949 and NSF-PHY-0134007.

\section{Discussion}

\noindent V.~Cavasinni (Univ. di Pisa/INFN):
{\it If the cosmic ray excess from the Galactic center is due 
to neutrons, would you expect also a neutrino flux coming from 
the neutron decays which could be measured by large volume 
neutrino detectors ?}\\
Yes.  Neutrons with energy less than EeV will decay in flight
and produce an antineutrino flux above TeV that could be detected
by the next generation neutrino telescopes such as IceCube.  The
expected event rate per year above 1\,TeV for IceCube was estimated
to be about 20.  See reference [51] for details.\\
\\
\noindent T.~Greenshaw (Liverpool University):
{\it Does the simultaneous observation of the fluorescence
and surface signals at Auger give additional information on 
the composition of UHE cosmic rays and, if so, do you have 
any preliminary results ?}\\
There is a wealth of information in the combined measurement, but 
studies are ongoing and there are no published results from Auger yet.  
The HiRes collaboration has published results based on the stereo
data.  See reference [52] for details.\\
\\
\noindent J.~Lefrancois (LAL, Orsay):
{\it You showed a plot with the Auger and AGASA energy spectra.
Can the different rate of almost a factor of three be explained 
by energy calibration ?}\\
The rate differences between AGASA, HiRes mono, and Auger indeed
suggest problems in the energy calibration.  Further evidence comes 
from the first analysis of the Auger spectrum.  Here, it has already 
been shown that there is a discrepancy in energies if the spectrum 
has its energy scale normalized to the fluorescence detector or if 
simulations are used to determine the shower energy from surface 
detector data alone.  This might indicate problems with the shower 
simulation code, the hadronic model, and the assumed composition of 
the cosmic ray flux.  The difference shows that many systematics
need to be addressed.

\end{document}